\documentclass[preprint]{revtex4}
\input epsf
\usepackage{amsmath,amssymb}
\usepackage{graphicx}

\draft

\begin{document}
\titlepage
\title{A Note on Energy-Momentum Conservation in Palatini
Formulation of L(R) Gravity}
\author{Peng Wang$^1$\footnote{E-mail: pewang@eyou.com},
Gilberto M. Kremer$^2$\footnote{E-mail: kremer@fisica.ufpr.br},
Daniele S. M. Alves$^2$\footnote{E-mail: dsma01@fisica.ufpr.br}
and Xin-He Meng$^{1,3}$\footnote{E-mail: xhm@physics.arizona.edu}
} \affiliation{1. Department of Physics, Nankai University,
Tianjin 300071, P.R.China \\ 2. Departamento de F\'\i sica,
Universidade Federal do Paran\'{a}, Caixa Postal 19044, 81531-990
Curitiba, Brazil \\ 3. Department of Physics, University of
Arizona, Tucson, AZ 85721 }

\begin{abstract}
By establishing that Palatini formulation of $L(R)$ gravity is
equivalent to $\omega=-3/2$ Brans-Dicke theory, we show that
energy-momentum tensor is covariantly conserved in this type of
modified gravity theory.
\end{abstract}

\maketitle

Nonlinear gravity theory of the $L(R)$ form has received a lot of
discussion due to their cosmological implications in early (see,
e.g., Ref.\cite{muk} for a review) and current Universe
\cite{Carroll}. Generally, $L(R)$ gravity has two inequivalent
formulations. One is the metric formulation in which the metric
tensor is viewed as the only dynamical variable in the
gravitational Lagrangian. The other one is the Palatini
formulation in which the metric and the connection are viewed as
independent dynamical variables.

Recently, Carroll \emph{et al.} \cite{Carroll} have proposed
adding a $1/R$ to the Einstein-Hilbert action to explain the
cosmic acceleration without dark energy (see Refs.\cite{chiba,
dolgov, odintsov-1/R, woodard} for subsequent discussions). In
Ref.\cite{Carroll}, the $1/R$ gravity is considered in the metric
formulation. However, there are at least two good motivations to
consider the Palatini formulation of $1/R$ gravity \cite{vollick,
wang, Flanagan, kremer, olmo, franca, ezawa} rather than the
metric formulation. First, for general $L(R)$, the field equation
in the metric formulation is fourth-order and this will cause
serious instability problem in $1/R$ gravity \cite{dolgov}. In
Palatini formulation, such instability problem is avoided
\cite{wang} (Note , however, that in generalizations of $1/R$
gravity which includes $R^2$ or $\ln R$ terms as were shown in
Refs.\cite{R2, lnR}, such instability is absent even in the metric
formulation). Second, Chiba \cite{chiba} has shown that metric
formulation of $1/R$ gravity is inconsistent with current solar
system experimental constraint on Brans-Dicke theory. In Palatini
formulation, now we believe that it is consistent with solar
system experiments \cite{Flanagan, olmo} (see also discussion at
the end of this note). Finally, it is also interesting to note
that the concept of minimal curvature in metric formulation of
$1/R$ gravity \cite{mini} extends straightforwardly to Palatini
formulation.

When considering the $L(R)$ gravity, an important question is
``whether energy-momentum tensor is covariantly conserved". In the
metric formulation of the $L(R)$ gravity, after some debates
\cite{metric}, it is finally established that energy-momentum
tensor is conserved for any form of $L(R)$ and from this follows a
`` generalized Bianchi Identity" \cite{magnano}.

Recently, in Ref.\cite{kremer}, by expanding the field equation to
first order, it is shown that in the Palatini formulation of the
$1/R$ gravity \cite{Carroll}, the energy-momentum tensor is
\emph{not} covariantly conserved. It is then important to ask
whether this is just the result of the perturbation expansion or a
true feature of the full theory. In this note, by establishing the
equivalence between the Palatini formulation of the $L(R)$ gravity
and the $\omega=-3/2$ Brans-Dicke theory, we show that
energy-momentum tensor is covariantly conserved in Palatini
formulation of $L(R)$ gravity.

When handled in Palatini formulation, one considers the action to
be a functional of the metric $\bar{g}_{\mu\nu}$ and a connection
$\hat{\bigtriangledown}_{\mu}$ which is another independent
variable besides the metric. The resulting modified gravity action
can be written as
\begin{equation}
S[\bar{g}_{\mu\nu},
\hat{\bigtriangledown}_{\mu}]=\frac{1}{2\kappa^2}\int
d^4x\sqrt{-\bar{g}}L(\hat R)+S_m\ ,\label{1}
\end{equation}
where we use the metric signature $\{-,+,+,+\}$, $\kappa^2=8\pi
G$, $\hat{R}_{\mu\nu}$ is the Ricci tensor of the connection
$\hat{\bigtriangledown}_{\mu}$,
$\hat{R}=\bar{g}^{\mu\nu}\hat{R}_{\mu\nu}$ and $S_m$ is the matter
action.

The field equations follow from varying the action (\ref{1}) with respect
to $\bar g_{\mu\nu}$,
\begin{equation}
L'(\hat R)\hat R_{\mu\nu}-{1\over2}\bar g_{\mu\nu}L(\hat R)=\kappa^2
T_{\mu\nu}^m\label{fieldeq}
\end{equation}
where $T_{\mu\nu}^m\equiv (-2/\sqrt{-\bar g})\delta S_m/\delta \bar
g^{\mu\nu}$ is the matter energy-momentum tensor. Also, by requiring
a stationary action with respect to variations of the connection
$\hat{\bigtriangledown}_{\mu}$, the latter turns out to be the Christoffel
symbol of the metric
$\hat g_{\mu\nu}=L'(\hat R)\bar g_{\mu\nu}$.

In Ref.\cite{Flanagan}, Flanagan has shown that the above action
is conformally equivalent to
\begin{equation}
\tilde S[\tilde g_{\mu\nu}, \Phi]=\int d^4x\sqrt{-\tilde
g}[\frac{\tilde R}{2\kappa^2}-V(\Phi)]\ ,\label{2}
\end{equation}
where
\begin{equation}
\bar g_{\mu\nu}=\exp(-2\kappa\Phi/\sqrt 6)\tilde g_{\mu\nu}\
.\label{3}
\end{equation}

Defining $\phi$ by the equation $\Phi=\sqrt 6/(2\kappa)\ln
L'(\phi)$, then $V(\Phi)$ is defined through $\phi$ by
\begin{equation}
V=\frac{\phi L'(\phi)-L(\phi)}{2\kappa^2L'(\phi)^2}\ .\label{4}
\end{equation}

Transforming the action (\ref{2}) back to the Jordan frame by
Eq.(\ref{3}) and make a field redefinition $\psi=\sqrt
6\exp(\kappa\Phi/\sqrt 6)/\kappa$, we find that action (\ref{1})
is equivalent to
\begin{equation}
S[\bar g_{\mu\nu}, \psi]=\int d^4x\sqrt{-\bar
g}[\frac{\psi^2}{6}\bar
R+\frac{1}{2}(\bar\nabla\psi)^2-V_\psi(\psi)]\ ,\label{5}
\end{equation}
where $V_\psi=\frac{\kappa^4}{36}\psi^4V(\frac{\sqrt
6}{\kappa}\ln(\frac{\kappa}{\sqrt 6}\psi))$. This is very similar
to the Induced gravity model with $\xi=1/3$ with the exception
that the sign of the kinetic term of the field $\psi$ is opposite
to the usual case.

Making a further field redefinition $\varphi=\psi^2/6$, the above
action can be rewritten exactly as the $\omega=-3/2$ Brans-Dicke
theory with a potential $V_\varphi(\varphi)$,
\begin{equation}
S[\bar g_{\mu\nu}, \varphi]=\int d^4x\sqrt{-\bar g}[\varphi\bar
R+\frac{3}{2\varphi}(\bar\nabla\varphi)^2-V_\varphi(\varphi)]\
,\label{6}
\end{equation}
where $V_\varphi=V_\psi(\sqrt{6\varphi})$.

As an example, for the $1/R$ gravity \cite{Carroll}, $L(\hat
R)=\hat R-\alpha^2/3\hat R$, the potential $V_\varphi$ is given by
\begin{equation}
V_\varphi=\frac{\alpha}{\sqrt{3}\kappa^2}(\kappa^2\varphi-1)^{1/2}\
.\label{7}
\end{equation}

Thus the field equations (\ref{fieldeq}) of $L(R)$ gravity in
Jordan frame can be rewritten in the form,
\begin{eqnarray}
\bar R_{\mu\nu}-{1\over2}\bar g_{\mu\nu}\bar
R=-{3\over2\varphi^2}[\bar\nabla_\mu\varphi\bar\nabla_\nu\varphi
-{1\over2}\bar g_{\mu\nu}\bar
\nabla_\lambda
\varphi\bar\nabla^\lambda\varphi]+{1\over\varphi}[\bar\nabla_\nu\bar
\nabla_\mu\varphi-\bar g_{\mu\nu}\bar\nabla^2\varphi]\cr+\bar
g_{\mu\nu}{V(\varphi)\over2\varphi}+{8\pi\over
\varphi}T_{\mu\nu}^m\ ,\label{field1}
\end{eqnarray}
and
\begin{equation}
-2V(\varphi)+\varphi V'(\varphi)=8\pi T^m\ .\label{field2}
\end{equation}

Now, by the well-known result that energy-momentum tensor is
covariantly conserved in Brans-Dicke theory (with a potential), or
can be checked explicitly using Eqs.(\ref{field1}) and
(\ref{field2}) (this would be extremely tedious using
Eq.(\ref{fieldeq}), but rather straightforward using the
equivalent formulation (\ref{field1}) and (\ref{field2}) ), we
conclude that energy-momentum tensor is covariantly conserved in
Palatini formulation of $L(R)$ gravity, i.e. $\bar\nabla^\nu
T_{\mu\nu}^m =0$. From this follows a ``generalized Bianchi
Identity":
\begin{equation}
\bar{\bigtriangledown}_{\mu}\left[L'(\hat R)\hat
R^{\mu\nu}-\frac{1}{2}L(\hat R)\bar{g}^{\mu\nu}\right]=0\
,\label{8}
\end{equation}
where $\hat R^{\mu\nu}$ and $\hat R$ are the Ricci tensor and
Ricci scalar constructed from the connection $\hat\Gamma$ which is
the Christoffel symbol with respect to the metric $L'(\hat
R)\bar{g}^{\mu\nu}$. This identity holds independently of the form
of $L(\hat R)$.

In the light of this, we wonder if in the expansion performed in
Ref.\cite{kremer} the terms neglected could provide a contribution
that would ratify the conservation of the energy-momentum tensor.
A more careful analysis reveals that, in fact, the higher order
derivatives of $L(\hat R)$ contribute with terms of order $\kappa
T/\alpha$ and therefore must be taken into account.

Let us consider the dust case where $p_{m}=0$ and $\rho_{m}=T$.
With the ansatz $\dot{T}=-\beta HT$, where, for the moment,
$\beta$ is kept undetermined, we obtain that the higher order
derivative terms yield the contribution:
\begin{equation}H\frac{\dot{L'}}{L'}\approx\beta\frac{\kappa T}{48}\qquad ,\qquad\frac{\ddot{L'}}{L'}\approx -\beta^{2}\frac{\kappa T}{48}\
.\label{9}
\end{equation}

Taking the contributions (\ref{9}) into account, the energy
density and pressure of the modified energy-momentum tensor of the
sources (as defined in Ref.\cite{kremer}) read:
\begin{equation}
\rho=\frac{\alpha}{4\kappa}+(11-\beta)\frac{\rho_{m}}{16}\
,\label{10}
\end{equation}
\begin{equation}
p=-\frac{\alpha}{4\kappa}+(\beta+1)(3-\beta)\frac{\rho_{m}}{48}\
.\label{11}
\end{equation}

Now, from $\dot{\rho}+3H(\rho+p)=0$, (\ref{10}) and (\ref{11}) it
follows that $\beta=3$ and
\begin{equation}
\dot{\rho}_{m}+3H\rho_{m}=0\ ,\label{12}
\end{equation}
and therefore the energy-momentum conservation is not spoiled.

Thus, by treating the $1/R$ gravity model in a perturbative form we
conclude that in the late Universe limit its dynamics reduces to
that of Standard GR with $\rho=\rho_{m}/2+\alpha/4\kappa$ and
$p=-\alpha/4\kappa$, that is, a dust-filled Universe with a
cosmological constant $\alpha/4\kappa$, where the coupling of the
geometry with matter is reduced by a 1/2-factor.

In sum, we established the equivalence between Palatini
formulation of $L(R)$ gravity and $\omega=-3/2$ Brans-Dicke theory
and using this to show that energy-momentum tensor is covariantly
conserved in this type of modified gravity theory. As a final
remark, it is interesting to notice that in Ref.\cite{chiba},
after establishing the equivalence between the metric formulation
of $1/R$ gravity and $\omega=0$ Brans-Dicke theory, we can find
that metric formulation of $1/R$ gravity is inconsistent with
current bound on the Brans-Dicke parameter $\omega$ from
measurements of the time delay using the Cassini spacecraft:
$\omega > 40,000$ \cite{omega}. Then whether the same conclusion
holds here? The answer is no: the bound on $\omega$ cannot be
applied to $\omega=-3/2$ Brans-Dicke theory. This can be seen from
Eq.(\ref{field2}). The Brans-Dicke field $\varphi$ in
$\omega=-3/2$ Brans-Dicke theory is not a propagating degree of
freedom, so $\omega=-3/2$ Brans-Dicke theory is actually not a
true scalar-tensor theory and thus bound on scalar-tensor theory
cannot be applied to it. This can also be seen in the fact that
the standard computation of the PPN parameters of Brans-Dicke
theory breaks down in the case of $\omega=-3/2$ \cite{will}. So
solar system experimental bound on PPN parameters cannot be
applied to this case. On the other hand, based on the discussion
in Refs.\cite{Flanagan, olmo}, we believe that $\omega=-3/2$
Brans-Dicke theory is actually consistent with current solar
system experiment.

\textbf{ACKNOWLEDGEMENT}

P.W. would like to thank Sergei D. Odintsov for helpful comment on
the manuscript and Clifford M. Will for helpful correspondence on
solar system constraint on Brans-Dicke theory. P.W. and X.H.M.
would like to thank Liu Zhao for helpful discussion on this topic.
X.H.M. would also like to express his thanks to the Physics
Department of UoA for its hospitality extended to him. This work
is supported partly by an ICSC-World Laboratory Scholarship, a
China NSF and Doctoral Foundation of National Education Ministry.

\end{document}